\documentclass[epj,final]{svjour}

\usepackage{graphicx}
\usepackage{amssymb}

\begin{document}

\title{Noise-induced reentrant  transition of the stochastic Duffing  oscillator}

\author{Kirone Mallick\inst{1} \and Philippe Marcq\inst{2}}     
\institute{Service de Physique Th\'eorique, Centre d'\'Etudes de Saclay,
 91191 Gif-sur-Yvette Cedex, France \\
\email{mallick@spht.saclay.cea.fr}\and 
Institut de Recherche sur les Ph\'enom\`enes Hors \'Equilibre,
Universit\'e de Provence,\\
49 rue Joliot-Curie, BP 146, 13384 Marseille Cedex 13, France\\
 \email{marcq@irphe.univ-mrs.fr}}

\date{March 1, 2004}

  \abstract{We derive the exact bifurcation diagram of 
   the Duffing  oscillator with parametric noise  thanks to
   the analytical study  of the  associated  Lyapunov exponent.
   When the fixed point is unstable 
   for the underlying deterministic dynamics,  we show that 
   the system undergoes a noise-induced reentrant  transition
   in a given range of parameters.
   The fixed point is stabilised when the amplitude of the noise 
 belongs to  a well-defined interval.
  Noisy oscillations are found  outside that range,
  {\it i.e.},  for both weaker and stronger noise.
   \PACS{
  {05.40.-a}{Fluctuation phenomena, random processes, noise, and 
  Brownian motion}   \and  {05.10.Gg}{Stochastic analysis methods 
  (Fokker-Planck, Langevin, etc.)}\and  {05.45.-a}{Nonlinear dynamics 
  and nonlinear dynamical systems}   
       } 
  }

 \maketitle 

 In a  classical calculation,  Kapitza  (1951) has shown  that 
the unstable upright position of an inverted pendulum can be stabilised
 if  its suspension axis undergoes sinusoidal vibrations
of high enough frequency \cite{landau}. 
 More generally, stabilisation can also be obtained 
with random forcing \cite{graham,luecke,landa}. In both cases, analytical
derivations of the stability limit are based on
perturbative approaches, {\it i.e.}, in 
 the limit of  small forcing  or noise amplitudes.

 In a recent work \cite{kirphil}, we studied the 
 Duffing oscillator with parametric white noise when  the  fixed point  of the 
 underlying deterministic equation
is stable: a purely noise-induced transition  \cite{lefever}
occurs when stochastic forcing is strong enough compared to
dissipation so as to `lift' the system away from the absolute 
minimum of the potential well. We showed, using a factorisation 
 argument, that the noise-induced
 transition occurs  precisely when  the Lyapunov 
exponent of the linearised stochastic equation changes sign.
 An analytical  calculation of the   Lyapunov exponent allowed us 
 to deduce  for all parameter values 
the   bifurcation diagram, that  was previously known  
 only in the  small noise  (perturbative)   limit. 
 The  relation between stochastic transitions  
 in a nonlinear Langevin equation 
 and the sign of  the Lyapunov  exponent
 is in fact  mathematically rigorous and has been proved
 under fairly general conditions  \cite{arnold}. In   \cite{kirphil}, 
 we also made the following  striking observation:  close 
 to the bifurcation, the averaged observables of the oscillator 
(energy,  amplitude square  and velocity square), as well as all 
their non-zero higher-order moments, scale linearly with the 
distance from the  threshold. This  multifractal behaviour may 
 provide  a  generic
 criterion to distinguish   noise-induced transitions from 
 ordinary deterministic bifurcations.

 In the present work, we  reformulate   the  stochastic 
 stabilisation   of an  unstable 
 fixed point of the underlying  deterministic system   within the framework
 of noise-induced transitions. We extend the analysis 
of  \cite{kirphil} and derive   the  full 
 phase diagram of  an  {\it inverted}  Duffing oscillator. 
  We show in particular that   a reentrant  transition
  occurs in this zero-dimensional system. (Reentrant
 transitions induced by noise  have been studied  in the more complex setting
 of spatially extended systems \cite{vandenbroeck}).  Moreover, the 
 inverted Duffing oscillator  also  exhibits 
 a multifractal behaviour close to the transition point.

 The dissipative stochastic system considered here 
  is defined by the equation:
\begin{equation}
\frac{\mathrm{d}^2 x}{\mathrm{d} t^2} + 
 \gamma  \frac{\mathrm{d} x}{\mathrm{d} t}
  +\sqrt{ {\mathcal D} } \xi(t) x =
- \frac { \partial{\mathcal U}}{\partial x}  \, ,
 \label{eqsto0}
\end{equation}
where $x(t)$ is the position of the oscillator at time $t$,
$\gamma $ the dissipation rate  
and $\xi(t)$ denotes a stochastic process. 
The confining, anharmonic  potential ${\mathcal U}(x)$ is defined as:
\begin{equation}
{\mathcal U}(x) = - \frac{1}{2} \mu  x^2 + \frac{1}{4} x^4 \,,
\label{potential}
\end{equation} 
where $\mu$ is a real parameter. Without noise, the corresponding
deterministic system undergoes a forward pitchfork bifurcation when
the origin becomes unstable as $\mu$ changes sign from negative to 
positive values.  In  \cite{kirphil}, we only considered   the case
 $\mu < 0$ and showed that  for
strong enough noise,  the origin becomes unstable. 
 Here, we study  the case 
$\mu > 0$ (`inverted'  Duffing oscillator) 
and   show that for  a finite range of positive values 
of $\mu$, a \emph{reentrant} transition is observed when
the noise amplitude is varied: the noisy oscillations obtained for weak and strong noise
are suppressed for noise of intermediate amplitude. Our results are non-perturbative:
  they are based on an exact calculation of the Lyapunov exponent,
performed for arbitrary parameter values,  $\xi(t)$  being  a 
Gaussian white noise process. 
Furthermore, our numerical simulations indicate that the phenomenology 
described above is unchanged for coloured Ornstein-Uhlenbeck  noise.

 We first rescale the   time variable 
 by taking the dissipative scale $\gamma^{-1}$ as the new time  unit.
 Equation~(\ref{eqsto0}) then becomes
\begin{equation}
\frac{\mathrm{d}^2 x}{\mathrm{d} t^2} + \frac{\mathrm{d} x}{\mathrm{d} t}
  - \alpha\; x  +  x^3  + \sqrt{\Delta}\;  \xi(t)  \; x = 0 \, , 
 \label{eqsto}
\end{equation}
where we have defined the dimensionless parameters
\begin{equation}
 \alpha =  \frac{\mu}{\gamma^2}
    \,\, \,\, \hbox{ and }  \Delta   =  
 \frac{ {\mathcal D} }{ \gamma^3 }  \, , 
  \label{defalpha}
\end{equation}
 and rescaled  the amplitude $x(t)$ by a factor  $\gamma$.
Linearising equation~(\ref{eqsto}) about the origin, 
we obtain the following stochastic differential equation:
\begin{equation}
\frac{\mathrm{d}^2 x}{\mathrm{d} t^2} + \frac{\mathrm{d} x}{\mathrm{d} t}
   - \alpha \; x  + \sqrt{\Delta}\;\xi(t)  \; x = 0 \, .
 \label{eqlin}
\end{equation}
The Lyapunov exponent of a stochastic dynamical system is generally defined 
as the long-time average of the local divergence rate from a given orbit
\cite{schimansky}. In the case discussed  here, deviations
from the (trivial) orbit  defined by the origin in phase space,  
$(x(t), \dot x(t)) = (0,0)$,  satisfy  equation (\ref{eqlin}).
 In practice, we use the (equivalent) definition for
 the (maximal)  Lyapunov exponent $\Lambda$ 
\begin{equation}
  \label{defnum}
  \Lambda = \lim_{t \to \infty} \frac{1}{2 \, t} \langle \log  x^2 \rangle \, ,
\end{equation}
where the brackets denote ensemble averaging.  Let
$z(t) = \dot x(t)/ x(t)$. From equation (\ref{eqlin}) 
we find that the new variable $z(t)$ obeys:
\begin{equation}
\dot z =  \alpha - z  - z^2 -   \sqrt{\Delta} \;  \xi(t) \,.
\label{eqz}
 \end{equation} 
The  Lyapunov exponent $\Lambda$ is equal to \cite{kirphil}
\begin{equation}
  \label{deflambda}
  \Lambda = \langle z \rangle_{\mathrm{stat}}
 = \int z \; P_{\mathrm{stat}}(z) \; \mathrm{d}z \,,
\end{equation}
where $P_{\mathrm{stat}}(z)$ is the stationary probability distribution
function (p.d.f.) of the variable $z$, solution of equation (\ref{eqz}).

\begin{figure}
\centerline{\includegraphics*[width=0.75\columnwidth]{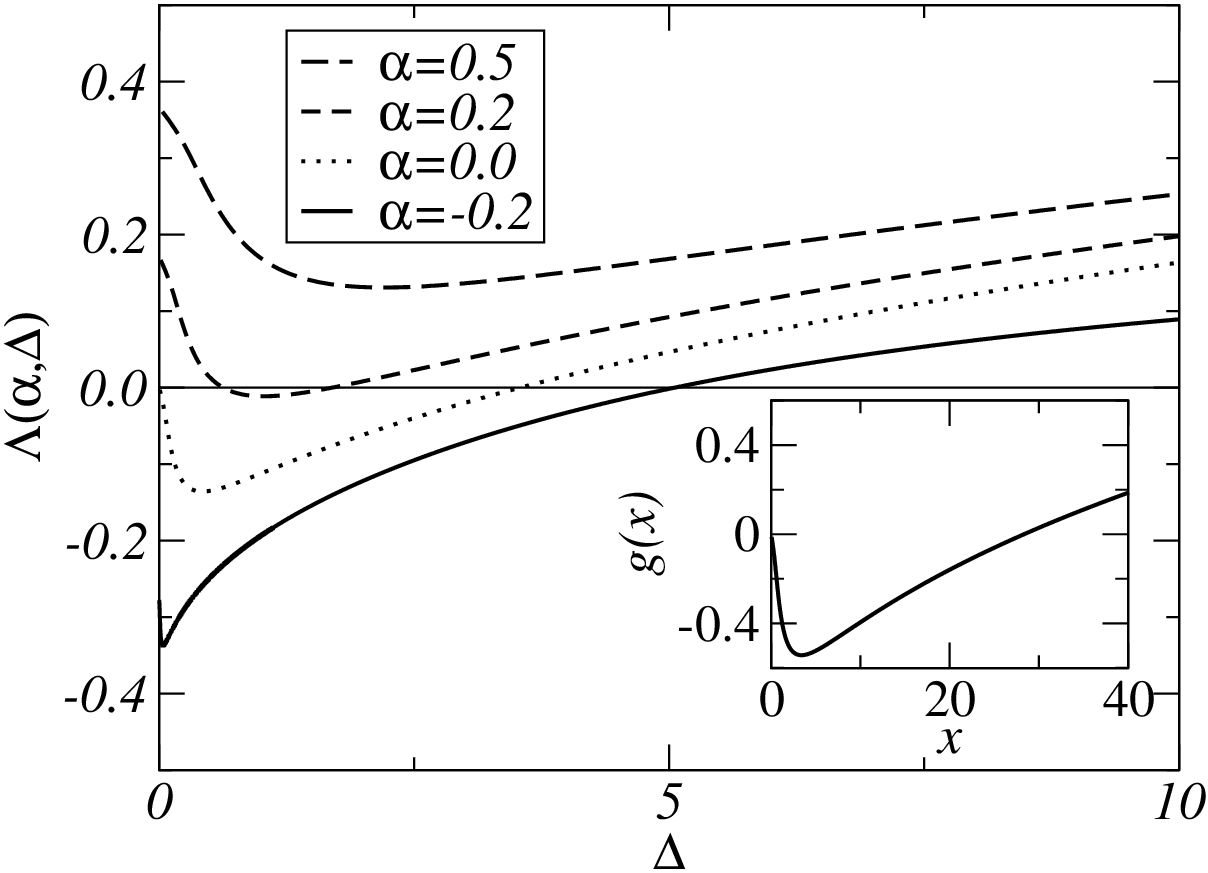}}

\caption{Lyapunov exponent $\Lambda$
of a linear, damped oscillator with parametric white noise. 
For different values of the control parameter $\alpha$, we plot
$\Lambda$ vs. the noise amplitude $\Delta$.
Inset: Graph of the function $g(x)$ (equation (\ref{defg})).}
\label{fig:lyap} 
\end{figure}

External fluctuations acting upon the oscillator will first be modeled
by Gaussian white noise $\xi(t)$  of zero mean value 
$\langle \xi(t)  \rangle=  0$ and of  unit amplitude:
\begin{equation}
  \label{whitenoise}
   \langle \xi(t) \xi(t') \rangle  =   \delta( t - t') \, .
\end{equation}
The stationary p.d.f. of $z$ can then be computed exactly,
by adapting a calculation presented in detail in \cite{kirphil}
(see also \cite{tessieri,imkeller}), where the Fokker-Planck equation
equivalent to equations (\ref{eqz})-(\ref{whitenoise}) is solved
in the stationary regime. We find:
\begin{equation}
 P_{\rm{stat}}(z) =  \frac{1}{N} \int_{-\infty}^z  \exp\left\{\frac{2}{\Delta} 
  (\Psi(z) - \Psi(y) ) \right\} \mathrm{d}y,
 \label{Pstatz}
\end{equation} 
where $\Psi(y) = \alpha y - \frac{1}{2} y^2 - \frac{1}{3} y^3$, and
$N$ is a normalisation constant. After some 
 calculations along the  lines of   Ref.~\cite{kirphil},
equation (\ref{deflambda}) leads to the following expression for  the 
Lyapunov exponent:
 \begin{equation}
  \Lambda(\alpha, \Delta)  = \frac{1}{2} \left\{  \frac {\int_0^{+\infty} 
\mathrm{d}u \; {\sqrt u} \;
  e^{ \frac{2}{\Delta} \left(  ( \alpha + \frac{1}{4} ) u
   - \frac{u^3}{12} \right) }  }
 {\int_0^{+\infty}  \frac{\mathrm{d}u}{\sqrt u}
  e^{\frac{2}{\Delta} \left(  ( \alpha + \frac{1}{4} ) u
   - \frac{u^3}{12} \right) } }
   -  1 \right\}   .      
\label{Lyapunov}
\end{equation} 
We  now examine the properties of 
$\Lambda$ as a function of $\alpha$ and $\Delta$. 
We shall limit the discussion to the range  $\alpha \ge -\frac{1}{4} $
(see \cite{kirphil} for a detailed study of the range $(- \infty, 0]$).
Let $\tilde \Delta = \Delta / (\alpha + \frac{1}{4})^{3/2}$. 
We  rewrite equation (\ref{Lyapunov}) as:
\begin{equation}
  \label{lyapg}
  \Lambda(\alpha, \Delta)  = \frac{1}{4} \; \sqrt{1 + 4 \alpha}
\; g(\tilde \Delta) + \frac{1}{2} \; \left( \sqrt{1 + 4 \alpha} - 1 \right),
\end{equation}
where the function $g(x)$ is defined for $x \ge 0$ as
\begin{equation}
  \label{defg}
  g(x) =  \frac {\int_0^{+\infty} \mathrm{d}u \; {\sqrt u} \;
  e^{ \frac{2}{x} \left(   u   - \frac{u^3}{12} \right) }  }
 {\int_0^{+\infty}  \frac{\mathrm{d}u}{\sqrt u}
  e^{\frac{2}{x} \left(  u   - \frac{u^3}{12} \right) } }   -  2 \,.
\end{equation}
The method of steepest descent  yields 
\begin{equation}
g(0) = 0 \, , \,\,\,\,  g'(0) = - \frac{1}{4} \, . 
 \label{eq:g0}
\end{equation}
The function $g(x)$   decreases  in  the interval
$[0, x_m]$,  with  an  absolute minimum
 $g(x_m) \simeq - 0.54$ at $x_m \simeq 3.40$, 
then $g(x)$  increases to infinity   over $[x_m, + \infty )$ 
(see the inset of Fig.~\ref{fig:lyap}). 
At fixed $\alpha$, the behaviour 
of $\Lambda(\alpha, \Delta)$ with respect to $\Delta$ is deduced from that
of $g(x)$ by translations and dilations along coordinate axes.
A few representative examples are drawn in 
Fig.~\ref{fig:lyap}:
$(i)$ $\alpha = -0.2$: $\Lambda$ changes sign once  
(noise-induced  bifurcation of the nonlinear system \cite{kirphil});
$(ii)$ $\alpha = 0.2$: $\Lambda$ changes sign twice 
(noise-induced reentrant transition of the nonlinear system);
$(iii)$ $\alpha = 0.5$: $\Lambda$ is positive for all $\Delta$ 
(no transition).

\begin{figure}
\centerline{\includegraphics*[width=0.75\columnwidth]{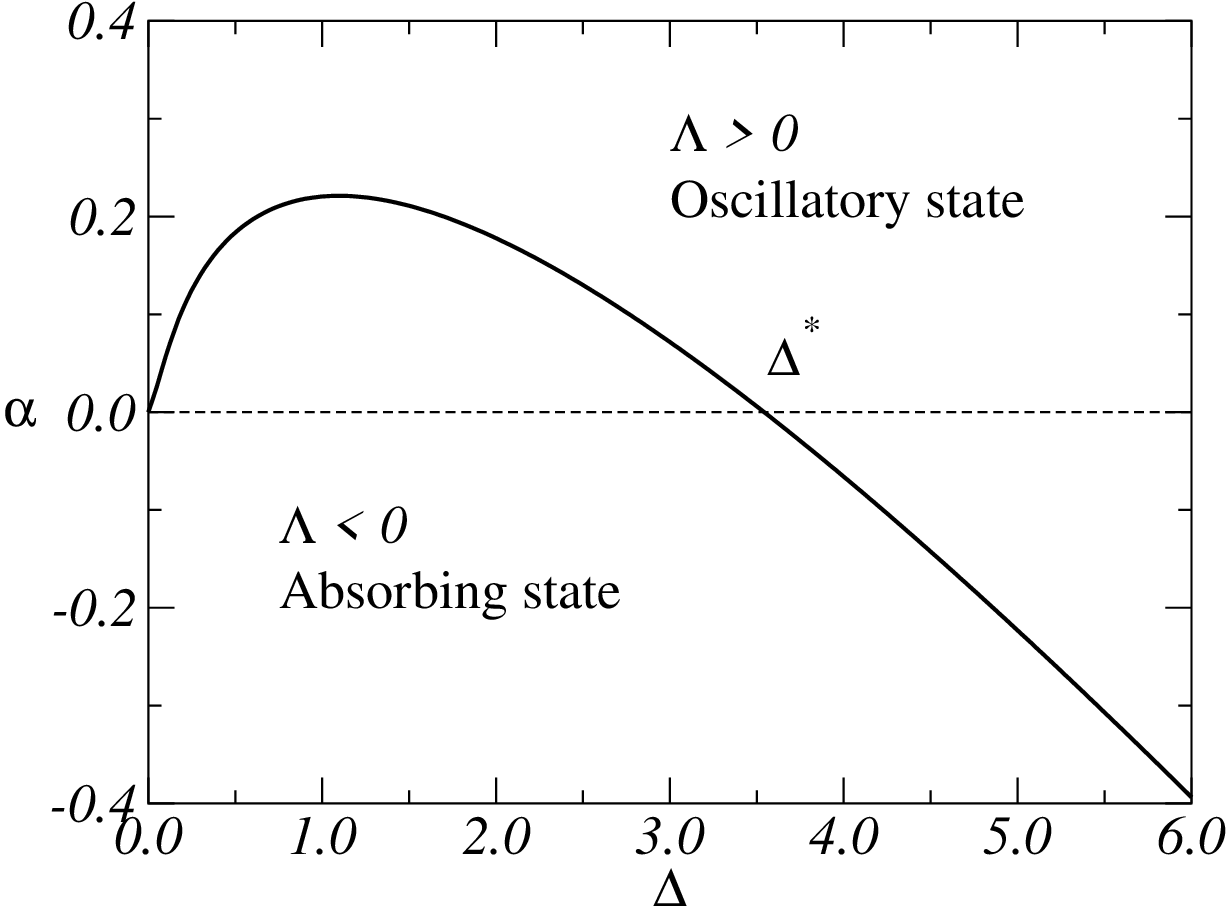}}

\caption{Bifurcation diagram of the Duffing oscillator with 
parametric white noise. The solid line is the locus in parameter space 
$(\alpha,\Delta)$ 
where $\Lambda(\alpha,\Delta) = 0$. The bifurcation line $\alpha = 0$ 
of the noiseless dynamical system is drawn for comparison (dotted line).}
\label{fig:diagbif} 
\end{figure}

   From this analysis, the bifurcation diagram of the nonlinear oscillator with 
parametric, Gaussian white noise (equations (\ref{eqsto} and 
\ref{whitenoise})) follows immediately. The origin is stable
(resp. unstable) when the Lyapunov exponent of the linearised
system is negative (resp. positive). The bifurcation line
$\alpha = \alpha_c(\Delta)$ 
 is defined  by  the equation 
\begin{equation}
\Lambda(\alpha_c(\Delta), \Delta) = 0 \,  . 
\end{equation}
 The transition
is best qualified as a stochastic Hopf bifurcation: the
bifurcated state displays noisy oscillations around
the origin, with a non-zero r.m.s. of the oscillator's position
and velocity.
For $\Delta \le \Delta^* \simeq 3.55$ (resp. $\Delta \ge \Delta^*$), 
the origin is stabilised (resp. destabilised) by the stochastic
forcing in the range $0 \le \alpha \le \alpha_c(\Delta)$
(resp. $\alpha_c(\Delta) \le \alpha \le 0$).
 The full bifurcation diagram of the  inverted stochastic 
 Duffing oscillator is displayed in Fig.~\ref{fig:diagbif}. 
 In  the weak noise limit $\Delta \to 0^+$, we obtain from Eq.~(\ref{eq:g0})
$\alpha_c(\Delta) \sim (-2 \; g'(0) \; \Delta) \sim  {\Delta}/{2}$,
in agreement with the result of the
 (perturbative) Poincar\'e-Linstedt expansion 
performed in \cite{luecke}.

We emphasise that the stability  of the origin of
 the   nonlinear random  dynamical system (\ref{eqsto}) \emph{cannot}
 be deduced from  a  stability  analysis of   finite-order moments of 
 the linearised system  \cite{lindenberg}: indeed 
second-order moments
of solutions of equation (\ref{eqlin}) with white noise forcing
are always unstable when 
$\alpha$ is positive. 
The proper indicator of the transition
of the nonlinear system is the Lyapunov exponent 
of solutions of the linearised equations.

\begin{figure}
\centerline{\includegraphics*[width=0.75\columnwidth]{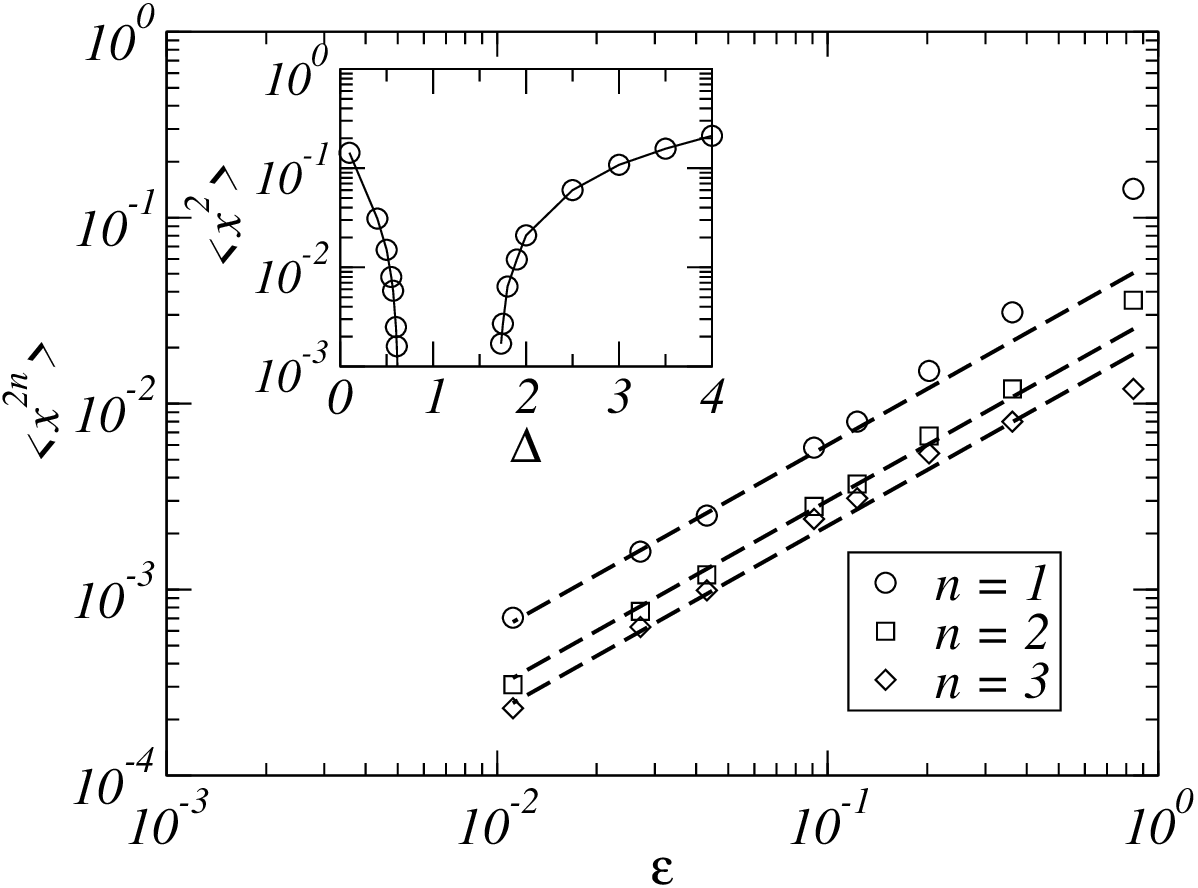}}
\caption{Reentrant transition of the noisy nonlinear oscillator
(\ref{eqsto}) subject to parametric 
white noise (\ref{whitenoise}), with a fixed parameter $\alpha = 0.2$. 
The numerical solutions of the equation $\Lambda(0.2,\Delta) = 0$
are $\Delta_1 \simeq 0.627$, $\Delta_2 \simeq 1.700$.
Even-order moments $\langle x^2 \rangle$, $\langle x^4 \rangle$
and $\langle x^6 \rangle$ are plotted versus the distance to threshold
$\epsilon = (\Delta_1 - \Delta)/\Delta_1$ (symbols). Dashed lines 
respecting a linear behaviour $\langle x^{2n} \rangle \propto \epsilon^1$ are 
drawn to guide the eye.  Linear scaling is also observed
for the second transition ($\Delta \gtrsim \Delta_2$; not shown).
Inset: the mean square position (measured in the stationary regime) 
is non-zero for $\Delta \in [0, \Delta_1[ \cup ]\Delta_2, + \infty[$. 
 }
\label{fig:reentrant} 
\end{figure}

 As an example, we show in Fig.~\ref{fig:reentrant} numerical 
evidence of a reentrant transition observed for $\alpha = 0.2$. 
The asymptotically stable state is the origin when the noise
amplitude belongs to the bounded interval $\Delta \in [\Delta_1, \Delta_2]$, 
where $\Delta_1$ and $\Delta_2$ are the two solutions of 
$\Lambda(\alpha = 0.2, \Delta) = 0$.
Noisy oscillations are found for both weaker ($\Delta < \Delta_1$)
and stronger ($\Delta > \Delta_2$) noise strengths.
As predicted in \cite{kirphil}, even-order moments
of the position and velocity scale linearly 
with the distance to threshold in the vicinity of these
stochastic bifurcations. (Note that odd-order moment are 
equal to zero by symmetry.)

Since the locus of the transition is determined by a dynamical property
of the linearised system, it cannot depend on the precise
functional form of the confining potential ${\mathcal U}(x)$ for large $x$.
For instance, the bifurcation line is unchanged when the confining
potential reads:
\begin{equation}
{\mathcal U}(x) = - \frac{1}{2} \alpha x^2 - \frac{1}{4} x^4
+ \frac{1}{6} x^6 \,,
\label{backward}
\end{equation} 
\emph{i.e.},  when the deterministic nonlinear system undergoes a 
\emph{backward} pitchfork bifurcation. We checked numerically that
such is indeed the case.

\begin{figure}
\centerline{\includegraphics*[width=0.75\columnwidth]{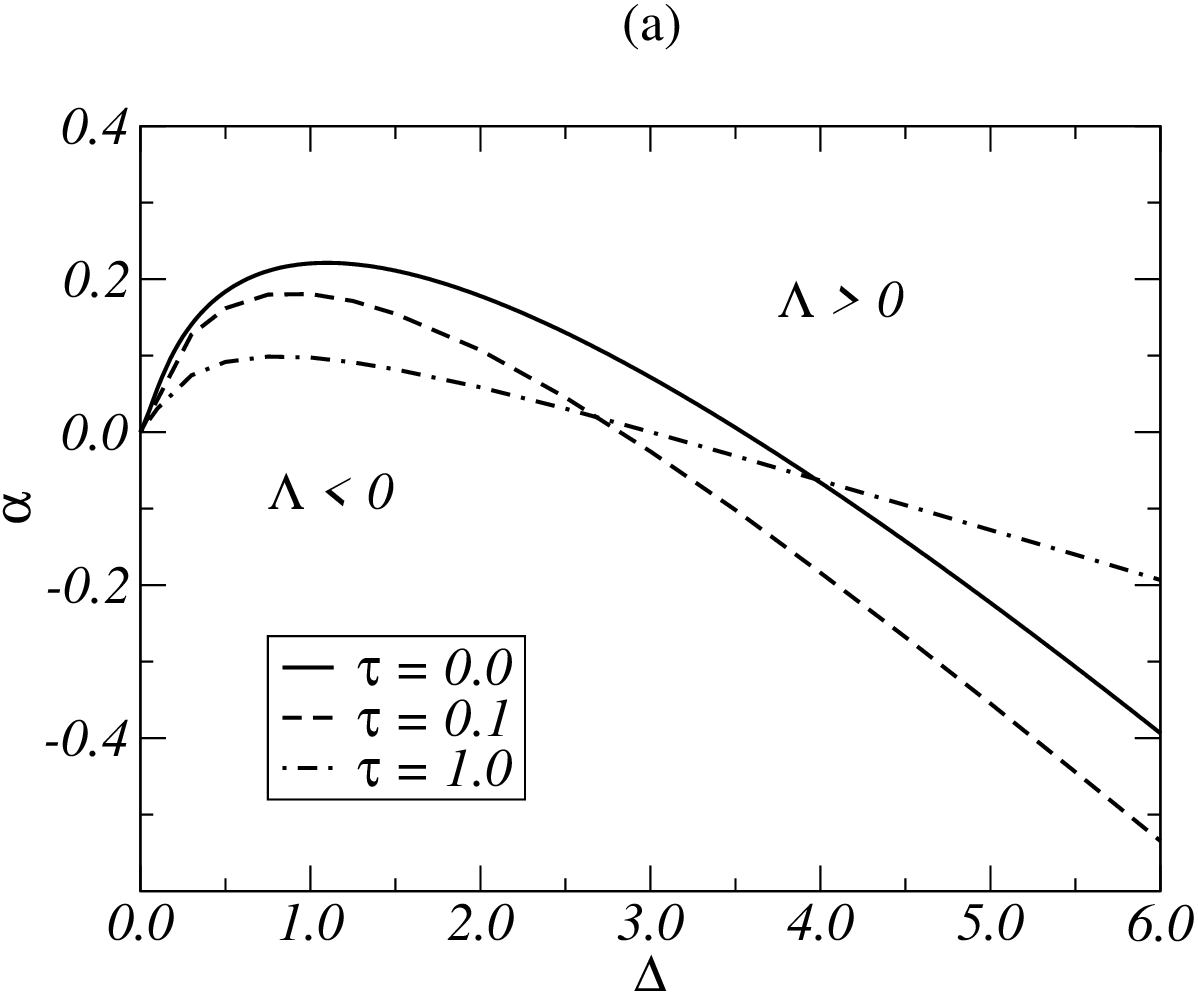}}

\bigskip
\centerline{\includegraphics*[width=0.75\columnwidth]{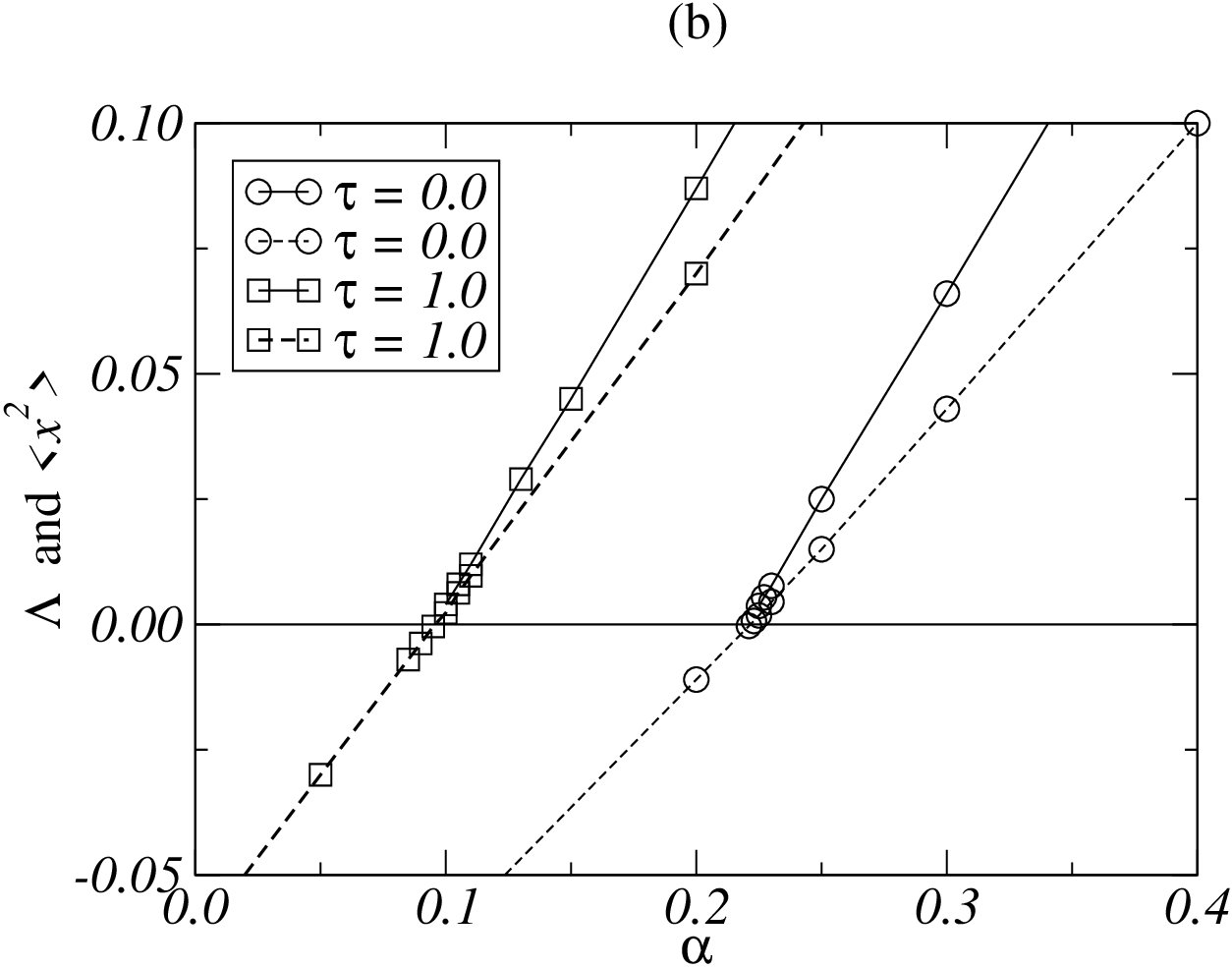}}
\caption{Parametric forcing with Ornstein-Uhlenbeck noise of correlation
time $\tau$. $(a)$ Bifurcation lines for $\tau = 0.1$ and $1.0$ are obtained
from numerical measurements of the Lyapunov exponent (\ref{defnum}).
For comparison, we draw the (analytic)
white noise line already given in Fig.~\ref{fig:diagbif}.
$(b)$  For $\tau = 1.0$ and $0.0$ (white noise), we plot on the 
same graph, for a fixed value $\Delta = 1.096$, and with respect 
to $\alpha$, 
the numerical values of the Lyapunov exponent of the linear system 
(symbols with dashed lines) and of the averaged
square position in the oscillatory regime of the nonlinear
system (symbols with solid lines). The bifurcation occurs
where the Lyapunov exponent changes sign. Note the linear 
dependence of $\langle x^2 \rangle$ on distance to threshold.}
\label{fig:OU} 
\end{figure}

Let us now turn to stochastic forcing by coloured noise. For definiteness,
we shall use an Ornstein-Uhlenbeck process $\xi(t)$ of correlation
time $\tau$, 
defined as the solution of the stochastic differential equation:
 \begin{equation} 
 \frac{{\textrm d} \xi(t)}{{\textrm d} t} = -\frac{1}{\tau} \xi(t) + 
\frac{1}{\tau} \eta(t) \, ,
  \label{OU}
\end{equation} 
where $\eta(t)$ denotes Gaussian white noise with zero mean and 
 unit amplitude 
  ($\langle \eta(t) \eta(t') \rangle  =    \, \delta( t - t')$).
The analytic expression of the stationary (marginal) p.d.f.
$P_{\mathrm{stat}}(z)$ associated with the linear system 
(\ref{eqz}-\ref{OU})
is not known: the Lyapunov exponent must  be evaluated numerically.
The line 
$\alpha = \alpha_c(\Delta)$ with $\Lambda(\alpha_c(\Delta), \Delta) = 0$ 
is drawn in Fig.~\ref{fig:OU}.a for $\tau = 0.1$ and $1.0$. 
(We checked that the same measurement protocol, when used with white noise 
forcing, yields data in agreement with the analytic result (\ref{Lyapunov}).)
Numerical simulations confirm that, with parametric coloured noise,
$(i)$ bifurcations of the nonlinear system 
(equations (\ref{eqsto}-\ref{OU}))  
occur where the Lyapunov exponent of the linear system changes sign; 
$(ii)$ averaged observables scale linearly with distance to threshold 
close to the bifurcation (see Fig.~\ref{fig:OU}.b).
In the weak noise limit, our data agrees with the prediction of \cite{luecke}:
$\alpha_c(\Delta) \sim  \Delta/(2 (1 + \tau))$.
For a  noise amplitude of order 1, the bifurcation line is qualitatively similar 
to that obtained with white noise. However, depending on the value of
$\Delta$, the value of the bifurcation point $\alpha_c(\Delta)$
is not necessarily   a monotonic function of $\tau$.

 We have shown here that noise can suppress oscillations
 by stabilizing a deterministically unstable fixed point. 
 Our study can be related to similar observations
 in the context of chaotic systems  where Lyapunov exponents
 also play a crucial role as indicators of transitions
 \cite{pikoskylett}. Indeed, the effective
 lowering of  Lyapunov exponents  by noise can induce synchronization
 in a pair of chaotic systems  \cite{banavar,sanchez}, or suppress
 chaos in  nonlinear oscillators \cite{rajasekhar}  or  in neural
 networks \cite{molgedey}. 

Our hope is that this work  will help to bridge the gap between
 the abstract  theory of random dynamical systems 
 \cite{arnold} and experimental
investigations of the stability of physical systems 
subject to random forcing \cite{fauve,petrelis}.
The behaviour of Lyapunov exponents of the linearised 
stochastic system (or ``sample stability'') is known
to explain the stability properties of a regular state
of electrohydrodynamic convection of nematic liquid crystals 
driven by multiplicative, dichotomous noise \cite{behn}.
A non-perturbative study of bifurcations of spatially-extended 
systems under stochastic forcing of arbitrary strength \cite{kramer}
remains a challenging open problem.

\end{document}